\documentstyle[12pt,fullpage]{article}
\begin{document}
\large
{\center {\Large SPIN-LATTICE RELAXATION IN ZERO-MAGNETIC FIELD 
INDUSED BY MOLECULAR REORIENTATIONS}

M.I.Mazhitov, Yu.A.Serebrennikov
\endcenter}
\vspace{1cm}
{\it The stochastic Liouville method is used to analyze the general problem of
spin-lattice relaxation in zero-field for molecules undergoing Markovian reorientations.}
\par
\vspace{1.2cm}
{\large {\bf I. Introduction}}
\par
  The resolution advantage of zero-field (ZF) studies for orientationatty
disordered materials is well known. In particular, the novel-pulsed ZF NMR and
NQR technique [I,2] offers an excellent approach to this problem since it
removes the orientational anisotropy which produces the broad high-field line
shapes in solids. In ZF NMR and NOR the signal comes from longitudinal nuclear
magnetization, i.e. the rank-I statisiical tensor.
 Time-domain ZF signals have also been observed using the method of perturbed
angular correlations of $\gamma$-ray  cascades [3]. From such experiments it is
possible to extract information concerning the zero-field spin-lattice
relaxation (ZF SLR) of rank $\geq 2$ statistical tensors. The corresponding response
function depends on the rates and microscopic details (in the slow-motional regime) 
of molecular reorientations which modulate the antisotropic part of the spin Hamiltonian.
\par
   Mathematical techniques have recently been developed which make analyses of
ZF NMR spectra in the complete tumbling regime feasible [4-6]. Our purpose here
is to extend this theory. We present a general formalism which enables us to
compute the response function of statistical tensors of arbitrary rank-$k$
irrespective of the models used to describe the Markovian molecular
reorientations. A compact expression for the correspondtng spectral function is
obtained, which is valid for the complete tumbling regime. To illustrate the use
of the theory we calculate the rank$-2$ perturbation coefficient of $\gamma$-ray cascades
arising from the quadrupole interaction of a spin$-1$ nucleus with an axially
symmetric electric field gradient. All the calculations in this paper are
confined  to situations that are macroscopically isotropic. There is continuity
with the formalism of the preceding articles [4-6] and intermediate results
derived there are assumed to have been looked at by the reader.
\par
\vspace{1.2cm}
{\large{\bf 2. Theory}}
\par
 In ZF the spin Hamiltonian for the problem is
\begin{equation}
\label{1}
\hat H(\Omega)=\sum_{q,p}(-1)^{p}\hat F_{2p}D^{2}_{q-p}(\Omega)A_{2q}
\end{equation}
\par
 Here $\hat F_{2p}$  is the $p$-component of a second-rank spin tensor operator in the
laboratory frame. $A_{2q}$  are components of a ZF splitting tenser expressed in the
molecular coordinate system(the principal axis frame) and $D_{q-p}^{2}(\Omega)$ are the Wigner
rotation matrices describing the transformation between the two frames. The
explicit form of $\hat F_{2p}$ and $A_{2q}$ will depend on the type of interaction.
\par
For sufficiently large molecules in dense media the stochastic reorientational 
process may be assumed to be Markovian. It then follows that an appropriate ensemble
average spin density operator $\hat\rho(\Omega,t)$ obeys the stochastic Liouville equation (SLE) [7,8]
\begin{equation}
\label{2}
\frac{\partial\hat\rho(\Omega,t)}{\partial t}=-iH^{x}\hat\rho(\Omega,t) + \hat{\hat L}_{\Omega}\hat\rho(\Omega,t)
\end{equation}
where $H^{x}\hat\rho=[\hat H,\hat\rho],  \hbar=1$ and $\hat{\hat L}_{\Omega}$ is the stationary Markovian operator describing 
the tumbling process. Eq.(\ref{2}) must be solved with the initial condition
\begin{equation}
\label{3}
\hat\rho(\Omega,0)=\phi(\Omega)\hat\rho(\Omega)=\frac{\hat\rho(0)}{8\pi^2}
\end{equation}
which takes into account the fact that for isotropic systems there is an equilibrium 
distribution of molecular orientations $\phi(\Omega)=\frac{1}{8\pi^{2}}$
\par
  The status of the spin ensemble can be discussed in terms of statistical tensors
$\rho^{(kp)}$(i.e. state multipole moments )
\begin{equation}
\label{4}
\hat\rho(\Omega,t)=\sum_{k,p}\rho^{(kp)}(\Omega,t)\hat T_{kp}(I)
\end{equation}
where the coefficient $\rho^{(kp)}(\Omega,t)$ and the irreducible polarization operator $\hat T_{kp}(I)$
[9] are given by 
$$
\rho^{(kp)}(\Omega,t)=Tr[\hat\rho(\Omega,t)\hat T_{kp}^{+}(I)],
$$
\begin{equation}
\label{5}
\hat T_{kp}(I)=\left(\frac{2k+1}{2I+1}\right)^{1/2}
\sum_{mm^{'}}C^{Im^{'}}_{Im\;kp}|Im^{'}\rangle \langle Im|
\end{equation}
Here $C^{Im^{'}}_{Im\;kp}$ is a Clebsch-Gordan coefficient. The corresponding response and spectral 
functions, $G^{kp}(t)$ and $\tilde G^{kp}(s)$, are obtained as averages over the equilibrium 
(isotropic) distribution:
\begin{equation}
\label{6a}
G^{kp}(t)=\int d\Omega\rho^{kp}(\Omega,t),
\end{equation}
\begin{equation}
\label{6b}
\tilde G^{kp}(s)=\int^{\infty}_{0}G^{kp}(t)\exp(-st)dt,
\end{equation}
where the tilde denotes Laplace transformation.
\par
 It thus follows from eqs. (\ref{5}) and (\ref{6a}),(\ref{6b}) that
\begin{equation}
\label{7}
G^{kp}(t)=Tr(\int d\Omega\hat D(\Omega)\hat\rho(\Omega)\hat D^{+}(\Omega)\sum_{q}(-1)^{p}D^{k}_{q-p}(\Omega))\hat T_{kq}(I)\equiv
\sum_{q}Tr[\hat\sigma^{kp}_{q}(t)\hat T_{kq}(I)]
\end{equation}
where $\hat D(\Omega)$ is the finite rotation operator. Following refs. [4,10], we multiply 
both sides of eq.(\ref{2}) by $\hat D(\Omega)$ on the left and by $(-1)^{p}\hat D^{+}(\Omega)
D^{k}_{q-p}(\Omega)$ on the right. 
In isotropic systems this procedure allows integration over $\Omega$ in the general form 
as reported in ref. [10]. Through a straightforward extension of the derivation
described in ref. [10] we obtain a compact differential kinetic equation
\begin{equation}
\label{8}
\dot{\hat\sigma}^{(kp)}_{q}(t)=-iH^{x}(0)\hat\sigma^{(kp)}_{q}(t)-\tau^{-1}\left(\hat\sigma^{(kp)}_{q}(t)-\sum_{q_{1}}\hat{\hat P}^{(k)}_{qq_{1}}\hat\sigma^{(kp)}_{q_{1}}(t)\right),
\end{equation}
where $\tau$ is the  mean lifetime between rotational jumps,
\begin{equation}
\label{9}
H^{x}(0)=\sum_{\mu}(-1)^{\mu}F^{x}_{2\mu}A_{2-\mu}
\end{equation} 
\begin{equation}
\label{10}
\hat{\hat P}^{(k)}_{qq_{1}}\hat\sigma^{(kp)}_{q_{1}}(t)=\int\hat D(\tilde\Omega)\hat\sigma^{(kp)}_{q_{1}}(t)\hat D^{+}(\tilde\Omega)D^{k}_{qq_{1}}(\tilde\Omega)f(\tilde\Omega)d\tilde\Omega,
\end{equation}
where $\tilde\Omega=\Omega-\Omega^{'}$ (see also refs. [4,10] ). The initial condition for eq. (\ref{8}) is obtained 
from eqs. (\ref{3}), (\ref{4}) and (\ref{7}):
\begin{equation}
\label{11}
\hat\sigma^{(kp)}_{q}(0)=\frac{\rho^{(kp)}(0)\hat T^{+}_{kq}}{2k+1}
\end{equation}
Formally eq. (\ref{8}) is similar to the impact equation which describes gas-phase relaxation. 
Reorientations may be classified as either "weak" or "strong" depending on the
angular jump, with its size set by the function $f(\Omega)$. The new formulation of the problem 
allows a solution irrespective of this circumstance in the general form.
\par 
  From that purpose let us re-express $\hat\sigma^{(kp)}_{q}$ in the form
\begin{equation}
\label{12}
\hat\sigma^{(kp)}_{q}(t)=\sum_{KQ}[\sigma^{(kp)}_{q}(t)]_{KQ}\hat T_{KQ}(I)
\end{equation}
It is easy to see that in this representation the response function can be written as
\begin{equation}
\label{13}
G^{(kp)}(t)=\sum_{q}(-1)^{q}[\sigma^{(kp)}_{q}(t)]_{k,-q}
\end{equation}
Then we have, using vector notation,
\begin{equation}
\label{14}
{\bf\dot X(t)= -(i\hat{\hat\Lambda}+\hat{\hat マ)X(t)}
\end{equation}
where the column vector {$\bf X(t)$} is constructed from the coefficients $\sigma^{(kp)}_{q}(t)]_{KQ}$.
The elements of the evolution, {$\bf\hat{\hat\Lambda}$} , and the motivational, {$\bf\hat{\hat マ=(1-\hat{\hat P}/\tau$} , operator matrices are [9]
$$
\hat{\hat\Lambda}^{qq_{1}}_{KQ\;K_{1}Q_{1}}=Tr(\hat H(0)[\hat T_{K_{1}Q_{1}}(I),\hat T^{+}_{KQ}(I)])\delta_{qq_{1}}=
$$
\begin{equation}
\label{15}
\sum_{K^{'}Q^{'}}(-1)^{Q}(K_{1}Q_{1};K-Q)^{K^{'}Q^{'}}Tr(\hat H(0)\hat T_{K_{'}Q_{'}}(I))\delta_{qq_{1}}
\end{equation} 
\begin{equation}
\label{16}
\hat{\hat マ^{qq_{1}}_{KQ\;K_{1}Q_{1}}=\sum_{LMN}W^{(L)}_{MN}C^{LM}_{kq\;KQ}C^{LN}_{kq_{1}\;K_{1}Q_{1}}
\end{equation}
where
$$
(K_{1}Q_{1};K-Q)^{K^{'}Q^{'}}\equiv(-1)^{2I+K^{'}}[(-1)^{K+K_{1}+K_{'}}-1][(2K+1)(2K_{1}+1)]^{1/2}\times
$$
\begin{equation}
\label{17}
\times C^{K^{'}Q^{'}}_{KQ\;K_{1}Q_{1}}\left\{\begin{array}{ccc}K&K_{1}&K^{'}\\I&I&I\end{array}\right\}
\end{equation} 
\begin{equation}
\label{18}
W^{(L)}_{MN}=(\delta_{MN}-A^{(L)}_{MN}),\quad A^{(L)}_{MN}=\int f(\tilde\Omega)D^{L}_{MN}(\tilde\Omega)d\tilde\Omega
\end{equation}
To derive eq.(\ref{16}) we have used the Clebsch-Gordan series for the product of Wigner 
matrices [ 9 ]. Eq.(\ref{14}) can be solved by Laplace transformation to give
\begin{equation}
\label{19}
{\bf\tilde X(}s{\bf)=\hat{\hat M}^{-1}(}s{\bf)X(}0{\bf)}
\end{equation}
where ${\bf M(}s{\bf)}=s{\bf\hat{\hat 1}}+i{\bf\hat{\hat\Lambda}}+{\bf\hat{\hat マ}$. It is easy to see from eqs. (\ref{11}) and (\ref{12}) that in this representation
\begin{equation}
\label{20}
{\bf X}(0)=[\sigma^{kp}_{q}(0)]_{KQ}=\left\{(-1)^{q}\frac{\rho^{(kp)}(0)}{2k+1}\delta_{Kk}\delta_{q,Q}\right\}
\end{equation}
Eq.(\ref{19}) is particularly suitable for numerical computation of the spectral function 
(\ref{6b}). The key step in the calculation is the inversion of matrix ${\bf\hat{\hat M}}$:        '
\begin{equation}
\label{21}
\tilde G^{(kp)}(s)=\sum_{qq_{1}}(-1)^{q+q_{1}}[{\bf\hat{\hat M}}^{-1}]^{qq_{1}}_{k -q, k -q_{1}}\frac{\rho^{(kp)}(0)}{2k+1}
\end{equation}
Since ${\bf\hat{\hat M}}$ has finite dimensions the inversion is readily achieved by standard techniques.
The result of eq.(21) provides a general recipe for calculating the response of
the rank$-k$ statistical tensor on Markovian molecular reorientations in ZF. The 
most severe restriction of the model is that the lattice is described only in terms 
of the orientationat degrees of freedom.
\par
  In the case osotropically rotating molecules, $f(\tilde\Omega)=f(cos(\tilde\beta))/4\pi^{2}$, from (\ref{16}) and (\ref{18}) we obtain
\begin{equation}
\label{22}
タ{qq_{1}}_{KQ\;K_{1}Q_{1}}=\sum_{L}\tau^{-1}_{\theta L}C^{L0}_{kq\;K-q}C^{L0}_{kq_{1}\;k-q_{1}}\delta_{KK_{1}}\delta_{q-Q}\delta_{q_{1}-Q_{1}}
\end{equation}
where $\tau^{-1}_{\theta L}=W{L}_{00}$ is the orientational relaxation time of the axial $L-$rank tensor. To illustrate 
the use of the theory we consider the case where the dominant anisotropic part
of the spin Hamiltonian is the axially symmetric quadrupote interaction [5]:
\begin{equation}
\label{23}
\hat H(0)=\sqrt{\frac{2}{3}}D_{Q}K_{I}\hat T_{20}(I)\equiv\sqrt{\frac{2}{3}}\frac{eQV_{zz}}{2I(2I-1)}K_{I}\hat T_{20}(I) 
\end{equation}
where $Q$ is the nuclear quadrupole moment,
\begin{equation}
\label{24}
K_{I}= (-1)^{2I}[\frac{1}{30}I(I+1)(4I^{2}-1)(2I+3)]^{1/2}
\end{equation}
As follows from (\ref{15}), (\ref{17}) and (23)
\begin{equation}
\label{25}
\hat{\hat\Lambda}^{qq_{1}}_{KQ\;K_{1}Q_{1}}=(-1)^{2I+K}[\frac{10}{3}(2K_{1}+1)]^{1/2}D_{Q}K_{I}C^{KQ}_{K_{1}Q\;20}
\left\{\begin{array}{ccc}2&K&K_{1}\\I&I&I\end{array}\right\}[(-1)^{K+K_{1}}-1]\delta_{qq_{1}}\delta_{QQ_{1}}
\end{equation}
As can be seen from (22) and (25) the components $[\sigma^{(kp)}_{q}]_K-q$ are uncoupled from the rest
of the vector ${\bf X}$ and the problem reduces to the inversion of the matrix ${\bf M}$ in $"Kq"$
subspace. It is convenient to calculate $\tilde G^{kp}(s)$ in the basis of eigenfunctions of the 
operator ${\bf\hat{\hat マ}$. In the $"Kq"$ subspace we have
\begin{equation}
\label{26}
[{\bf\hat{\hat U}}^{-1}{\bf\hat{\hat マ\hat{\hat U}}]_{Kn\;K_{1}n_{1}}=タ{'}_{Kn\; K_{1}n_{1}}=\gamma_{n}\delta_{KK_{1}}\delta_{nn_{1}},\quad
\hat{\hat U}_{Kq\; Kn}=C^{n0}_{kq\;K-q}
\end{equation}
$\hat{\hat U}$ is the unitary matrix which makes the submatrix $[タ{'}]^{qq_{1}}_{K-q\; K-q_{1}}$ in (22) diagonal, 
$\gamma_{n}=\tau^{-1}_{\theta n}$  denotes the eigenvalues. It is easy to see that in this representation eqs. (\ref{13}) and (21) give
\begin{equation}
\label{27}
\tilde G^{(kp)}(s)=\left[\frac{1}{s{\bf\hat{\hat 1}}+i{\bf\hat{\hat\Lambda}^{'}}+{\bf\hat{\hat マ}}\right]_{k0,k0}\rho^{(kp)}(0),
\end{equation}
where                           .
$$
\hat{\hat\Lambda}^{'}_{Kn\; K_{1}n_{1}}=(-1)^{2I+k}2D_{Q}K_{I}\left[\sqrt{\frac{2}{3}}(2K+1)(2K_{1}+1)(2n+1)(2N_{1}+1)\right]^{1/2}\times
$$
\begin{equation}
\label{28}
\times C^{20}_{n0\;n_{1}0}\left\{\begin{array}{ccc}K&K_{1}&2\\I&I&I\end{array}\right\}\left\{\begin{array}{ccc}K&K_{1}&2\\n_{1}&n&k\end{array}\right\} (\delta_{K K_{1}+1} + \delta_{K K_{1}-1})
\end{equation}
Consequently only one element of the inverted matrix ${\bf(\hat{\hat M})}^{-1}$ is needed to calculate the spectral function.
\par
\vspace{1.2cm}
{\large{\bf 3. Discussion}}
\par
  In the fast motional limit, $D_{Q}\tau_{\theta 2} \ll 1$ , taking into account (26)-(28) to second order in 
perturbation theory, we have
$$
G^{(kp)}(s)=(s+\lambda_{k})^{-1}\rho^{(kp)}(0),\quad G^{(kp)}(t)=\exp[-\lambda_{k}t)\rho^{(kp)}(0)
$$
where
$$
\lambda_{k}=\frac{3}{80}(eQV_{zz})^{2}\tau_{\theta 2}\frac{k(k+1)[4I(I+1)-k(k+1)-1]}{I^{2}(2I-1)^{2}}
$$
in agreement with Abragam and Pound[11]

For a simple illustration of the formalism introduced in section 2 we consider a
case with $I=1$. From (26)- (28) we obtain
\begin{equation}
\label{29}
\tilde G^{1p}(s)=\frac{(s+\tau^{-1}_{\theta2})^{2} + \frac{1}{3}D^{2}_{Q}}{s[(s+\tau^{-1}_{\theta2})^{2}+D^{2}_{Q}]+\frac{2}{3}D^{2}_{Q}\tau^{-1}_{\theta2}}\rho^{1p}(0)
\end{equation}
\begin{equation}
\label{30}
\tilde G^{2p}(s)=\rho^{2p}(0)\frac{N_{1}(s)}{sN_{1}(s) + N_{2}(s)},
\end{equation}
where 
$$
N_{1}(s)=\frac{1}{7}D^{2}_{Q}(s+\tau^{-1}_{\theta4}) + \frac{16}{35}D^{2}_{Q}(s+\tau^{-1}_{\theta2})+(s+\tau^{-1}_{\theta2})^{2}(s+\tau^{-1}_{\theta4})
$$ $$
N_{2}(s)=\frac{2}{5}D^{2}_{Q}(s+\tau^{-1}_{\theta2})(s+\tau^{-1}_{\theta4})
$$
Eq. (\ref{29}) is identical to there suh of ref. [5](eq.(24)) which describes the ZF
NMR spectral function [1,2]. Ordinary NMR coils can only detect rank-1 tensors.
However, by experimental observaiion of $\gamma$-ray cascades, it is possible to
extract information concerning the relaxation of $k \ge 2$  statistical tensors. In
particular, the measurer anisotropy [3, 11, 12] is proportional to $G^{k0}(t)$. Sometimes
it is convenient to observe the average correlation of all decays:
$$G^{k0}(\infty)=\tau^{-1}_{N}\int^{\infty}_{0}G^{k0}(t)\exp(-\frac{t}{\tau_{N}}),$$
which is just the Laplace transformation at $s=\tau^{-1}_{N}$, where $\tau_{N}$ is the mean nuclear
lifetime.
\par
 The spectral function (30) depends on $\tau_{\theta2}$  and $\tau_{\theta4}$. In the "strong collision"
model $f(\Omega)=\frac{1}{8\pi^{2}}$ and $\tau_{\theta2}=\tau_{\theta4}=\tau$. 
In contrast, $\tau^{-1}_{\theta n}=n(n+1)D_{r}$ under Debye orientational diffusion 
with coefficient $D_{r}$. Thus in the slow tumbling regime the precise form of the
angular correlation depends on dynamical details of the motion.
\par
 It is interesting to compare our exact result (30) with the approximate
analytical solution of the problem which has been obtained by Linden-Bell [12].
It is easy to see (by Laplace transformation of thc corresponding
expressions for $G^{20}(t)$ which have been given in ref.[12] that in the fast
motional regime our results coincide. However, her approximation is not
sufficient to obtain good quantitative agreement wich  eq. (30) in
the slow-motional limit, when $D_{Q}\tau_{\theta2} \simeq 1$.
\newpage
\vspace{1.5cm}
{\large References}
\par
\begin{enumerate}
\item D. Weitkamp, A. Sielecki, D. Zax, K. Ziim and A. Pines, Phys.Rev.Letters
50 ( 1983) 1897.
\item A. Thayer and A. Pines, Accounts Chem. Res. 20 (1987) 47.
\item R.M. Steffen and H. Frauenfelder, in: Perturbed angular correlations, eds.
E. Karlson, E. Matthias and K. Siegbahn (North-Holland, Amsterdam, 1964) p. 3.
\item Yu.A.Serebrennikov, Chem.Phys. 112 ( 1987) 253.
\item Yu.A.Serebrennikov, Chem,Phys.Letters 137 (1987) 183.
\item Yu.n.Serebrennikov, M.I. Majitov and Z.M. Muldakhmetov, Chum. Phys. 121 
(1988) 307.
\item A.I.Bershtein and Yu.S.Oseledchik, Soviet Phys.JETP 51 (1966) 1072.
\item R.Kubo, Advan.Chem.Phys. 16 (1969) 101.
\item D.A. Varshalovich, A.N. Moskalev and V.K. Khersonsky, Quantum theory
of angular moment (Nauka, Moscow, 1975).
\item Yu.A Serebrennikov, S.I. Temkin and A.I. Burshtein, Chem.Phys. ( 1983) 31
\item A.Abragam and R.V.Pound, Phys.Rev. 92 ( 1953) 943.
\item R.Lynden-Bell, Mol.Phys. 22 (1971) 837.             ,

\end{enumerate}

\end{document}